# Automatic Authorities: Power and AI[1]

Seth Lazar, Australian National University

> *Man, a child in understanding of himself, has placed in his hands physical tools of incalculable power. He plays with them like a child, and whether they work harm or good is largely a matter of accident. The instrumentality becomes a master and works fatally as if possessed of a will of its own— not because it has a will but because man has not. (Dewey, 2016 (1926): 197)*

**Introduction**

As rapid advances in Artificial Intelligence and the rise of some of history's most potent corporations meet the diminished neoliberal state, people are increasingly subject to power exercised by means of automated systems. Machine learning, big data, and related computational technologies now underpin vital government services from criminal justice to tax auditing, public health to social services, immigration to defence (Citron, 2008; Calo and Citron, 2020; Engstrom et al., 2020). Google and Amazon connect consumers and producers in new algorithmic markets (Nadler and Cicilline, 2020). Google's search algorithm—and possibly in the near future OpenAI's GPT-4 or another large language model—determines, for many, how they find out about everything from how to vote to where to get vaccinated. Meta, Twitter, TikTok, Google and others algorithmically decide whose speech is amplified, reduced, or restricted (Vaidhyanathan, 2011; Pasquale, 2015; Gillespie, 2018; Suzor, 2019). And a new wave of products based on rapid advances in Large Language Models (LLMs) have the potential to further transform our economic and political lives.

*Automatic Authorities* are automated computational systems used to exercise power over us by substantially determining what we may know, what we may have, and what our options will be. In response to their rise, scholars working on the societal impacts of AI and related technologies have advocated shifting attention from the question of how to make AI systems beneficial or fair towards a critical analysis of these new power relations (Boyd and Holton, 2018; Bucher, 2018; Liu, 2018; Nemitz, 2018; Susskind, 2018; Cohen, 2019; Crawford, 2021; Véliz, 2021). But what normative lessons should we draw from these analyses? Power is everywhere, and is not necessarily bad. Why should the fact that Automatic Authorities are used to exercise power (or perhaps even exercise power themselves) cause moral concern?

---

[1] This chapter is based on, and substantially revises, my 'Power and AI: Nature and Justification', in the *Oxford Handbook of AI Governance* (Justin Bullock et al., eds). My thanks to the publisher for their permission to use this material.

This chapter introduces the basic philosophical materials with which to formulate these questions, and offers some preliminary answers. It starts by pinning down the slippery concept of power, focusing on the ability that some agents have to shape others' lives. It then explores how AI enables and intensifies the exercise of power so understood, and sketches three normative problems with power and three complementary ways to solve those problems. First, however, it is important to pin down AI itself.

AI is a set of technologies enabling computer systems to perform functions like (at the highest level of abstraction) making inferences from data, generating text and images, optimizing for the satisfaction of goals within constraints, and learning from past behaviour (Russell and Norvig, 2016). Machine learning (ML) is a field of AI using algorithms, statistical models, and recursive simulation to detect patterns in massive datasets, enabling computers to perform these functions without following explicit instructions. Advances in computational power and an explosion in the availability of training data, as well as some non-trivial theoretical discoveries, have led to progress in ML that has made AI systems startlingly effective at these tasks, enabling them to play a pervasive role in our lives. Recent advances in extremely large language models (LLMs, see Vaswani et al., 2017; sometimes called Foundation Models, Bommasani et al., 2021), using extraordinary amounts of computation and data, promise yet further significant practical advances, with new research papers being released almost daily identifying previously unrecognised latent capabilities of these models, especially when fine-tuned for particular practical applications (Singhal et al., 2022; Wei et al., 2022; impressive as these preliminary results are, however, it is yet to be seen whether results from LLMs will be reliable enough for use in high-stakes contexts).

Definitions are needed so that we don't talk past each other; but little often turns on whether some particular computational system is properly described as 'AI'. If it involves computation and some measure of automation, then many of the same ethical and political issues arise whether it's ultimately powered by deep neural nets, by simple logistical regression, or even by old-fashioned symbolic reasoning. Algorithmic tools like blockchain, hash-matching, and (non-ML) search often raise just the same normative questions as does AI. As one way to see this, consider early papers by Roger Clarke, Helen Nissenbaum, and Danielle Citron, all of which long antedate contemporary progress in ML, but which describe moral and political problems of computing systems—'dataveillance', diffuse accountability, power without due process—that remain entirely relevant today (Clarke, 1988; Nissenbaum, 1996; Citron, 2008).

**Power**

The next task is to fix what we mean by power. This concept is widely invoked in the literature on AI, including now in papers making fundamental technical advances, as they explore the ethical implications of their work (e.g. Singhal *et al.*, 2022). But scholars and commentators often just assume that we all mean the same thing by the concept, and that its normative significance is equally transparent. It is not, however, that simple. There are many different ways to understand power—we need to decide which are most salient.

The first key distinction is between 'power to' and 'power over' (Dowding, 2012; Pansardi, 2012). On the first approach, power is a resource distributed around society—some people, or social groups, have more power than others, and we can think of it in terms of what individuals or groups have power to do (Jenkins, 2009). This can be understood very expansively—so that power to is really just a way of describing ability (Morriss, 2002); or more narrowly, for example by considering different social groups' ability to affect collective decision-making (Goldman, 1972). We might say, for example, that wealthier voters have disproportionate power to influence political outcomes compared to low-income voters.

On the second approach, power over describes a social relation whereby some entity has power over some other entity. For Alice to have power over Bob means, minimally, that Alice can get Bob to do things that Bob wouldn't otherwise do—that Alice can control Bob (Dahl, 1957; Barry, 1974). This can operate in different ways. Alice can shape Bob's behaviour by: directly impacting Bob's prospects—what makes his life go better or worse; by altering Bob's options; and by influencing Bob's beliefs and desires. Alice breaking Bob's arm is an exercise of power over him, just in its own right. But suppose Alice works for a firm that competes with Bob's, and she wants him to share corporate secrets with her, which he does not want to do. If she tells him she'll have his arm broken if he does not comply, then she's changing his options—removing the option to stay silent without suffering material harm. Or else if she hacks into his computer so that his emails are automatically uploaded to a server where she can read them—again she's changing his options. She can also exercise power over him more indirectly: by sharing evidence with him that shows how his revealing corporate secrets would actually benefit society at large, for example, thereby changing both his beliefs and desires.

Most scholars who emphasise the importance of thinking about AI and power are thinking about power over. But AI has significant implications for power to, as well. Think, for example, of the ways in which AI, robotics, and related technologies increase our capacities to get things done. The power to that AI generates is unevenly distributed—it empowers some more than others. Most obviously, it empowers those who have access to it, which is in general those who are already advantaged. In other words, AI increases the power to bring about desired results of those who are already privileged. This is of obvious moral importance. One potential promise of LLMs—if their emergent abilities prove sufficiently robust—is to radically empower people who currently cannot afford access to professional expertise. Diffusion models already empower those with limited artistic skill to generate aesthetically pleasing images.

We ought not disregard power to. However, understood in this way, we *can* assimilate it to other concepts used in a theory of distributive justice—which concerns how the benefits and burdens of being part of society should be distributed. In general, increasing people's power to do non-harmful things is an unalloyed good—the big questions concern how to make sure one doesn't also increase their power to do harmful things, and to make sure that these new abilities are fairly distributed. For example, as AI empowers the unskilled to create attractive illustrations, how should society restructure its creative economy so that skilled artists are able to make a living, without just relying on AI? Power over, as will be shown below, raises somewhat distinctive normative questions, with particular urgency in the age of AI. For that reason this chapter focuses on power over.

Power over is a two-place relation: A has power over B. The first task is to populate those two places. The simplest example involves one person having power over another person—Alice over Bob. Power can also be exercised over other entities, such as animals, but interpersonal power is the focus here.

Individuals, groups, and aggregates can have power over individuals, groups, or aggregates. 'Groups' are sets of individuals with some internal structure—for example, a set of rules for membership or for decision-making. A company like Meta is a group. An 'aggregate' is just a set of individuals or groups without such an internal structure. The users of Facebook, for example, are an aggregate.

On a different understanding of power over, people can not only be subject to the power of agents but also to that of social structures (Dowding, 2008; Haslanger, 2012). Social structures are (roughly) networks of roles, relationships, incentives, norms, and cultural schemas (widely shared sets of beliefs and desires), which can be populated or observed by different people at different times, which are generally the emergent result of human interaction over time, and which reliably pattern outcomes for people who are within or affected by them (Haslanger, 2016; Ritchie, 2020). The presence of a particular social structure increases the probability that people who meet a particular description will experience a particular kind of outcome. Indeed, social structures can clearly have effects on people which, if they were traceable to some individual's decision, would lead us to say that that individual has power over those people. So, do social structures have power over us?

This question has exercised social theorists for decades, and we cannot settle it here. The concept of power is arguably linguistically appropriate for both cases. However, for present purposes power is most usefully understood as a social relation between people, or at least agents, within society (call this agential power). Social structures often shape that relation, by giving some people power over others, but all power is ultimately exercised by agents. What's more, the distributional effects of social structures on people's prospects and opportunities can be adequately described with other concepts, whereas the social and agential relation of power over cannot be described in any other way.

It is important to remember, however, that a theory of agential power definitely demands a theory of the political affordances of social structures, which often clearly determine whether and to what extent one person has power over another (Young, 1990; Haslanger, 2012). Think, for example, of the way in which social structures of law, policing and criminal justice in the United States give White Americans power over Black Americans. And further research is necessary on the ways in which social structures are encoded into ML models—especially LLMs—in ways that potentially give some people power over others.

Stakes matter. How should we measure power, so we can tell how morally significant or insignificant it might be? There are at least three important dimensions.

First: the *degree* of A's power over B. How substantial and robust are the effects that A can visit on B? Does A have power over whether B lives or dies? Or is it just a matter of determining what advertisements B will see when she browses the internet? Has A brainwashed B, so as to entirely determine B's beliefs and desires? Or can A do no more than conceal or display some salient facts? As well as actual harms and benefits, we should consider the *risks* to which A exposes B.

Can we be subject to the power of only human agents? Could we be subject to the power of computing systems? If an automated system can effect a change in the world that would amount to an exercise of power if done by a person, that certainly *looks* like agential power. But if the system is simply deploying a set of pre-programmed rules, it might be only a tool for the exercise of power, rather than itself exercising power—we're *really* subject to the people who put the system in place over us. More generally, if the system is subject to the effective control of those who design or deploy it, then they are the ones with power. But if the system's programming instead derives from its own learning, and if it operates at a speed and scale beyond the effective control of those who design and deploy it, then perhaps the computing system itself exercises power. As the field of Machine Learning advances, it will become less and less controversial to say that AI systems themselves are exercising power—noting that the conditions of agency required to say that a system exercises power are quite different from the conditions appropriate to saying that a system is, for example, a *moral* agent, as well as from assertions that the system is sentient, conscious, or anything more metaphysically surprising. Roughly speaking, if A can substantially affect B, A has the ability to do otherwise, and A is not under the effective control of some other agent, then A very likely has power over B. Dialogue agents based on LLMs, like ChatGPT and Claude, seem pretty clear cases of Automatic Authorities that exercise power over their users, for example.

Importantly, sometimes A has the ability to affect *very many* people by a small amount. This might constitute very little power over any individual B, but a significant degree of power over 'the Bs'. For example, suppose the Bs are an evenly divided population in a first-past-the-post democracy, and A can influence each person by a fraction of a per cent to accord with A's political views. This small individual effect could easily translate into a very significant aggregate effect.

The *scope* of A's power over B is also important—the proportion of B's life over which A has influence, the range of different direct effects, and different choices that A can indirectly affect. If A has significant power over B, but only within a narrow range of B's choices, then that amounts to a less significant instance of power than if the same degree of power is realised over a broader swathe of B's life.

Lastly the concentration of power is important—this takes us beyond the power relations between A and B, and asks how many people A has power over (to what degree and scope in each case). Other things equal, power is more concerning when it is more concentrated.

**How AI is Used to Exercise Power**

With this working understanding of power in hand, we can start to explore how we are already subject to the power of Automatic Authorities in many different domains of our lives, and how this is only likely to increase in the future. I'll consider each of the three main modalities for the exercise of power: intervening on people's interests; shaping their options; and shaping their beliefs and desires.

Start with effects on the subject's interests. AI systems are frequently used to support the allocation of resources within a population. This amounts to the exercise of power by the decision-maker (the individuals or organizations making use of the AI decision-support tool) over those who either benefit or don't from that decision. In the public sector, think of the allocation of healthcare, housing, social welfare resources, or algorithmic allocation of visas (Eubanks, 2017). In the private sector, think of insurance pricing, decisions over whether a loan is granted, or automated systems that determine what products, services, and content you are exposed to online (Gorwa et al., 2020).

What AI gives, AI can also take away: AI is used to allocate harms, as well as to directly harm people. The use of AI in the criminal justice system to inform decisions over pre-trial detention, sentencing, and parole is an obvious example where AI is used to exercise one of the most profound and serious powers that the state holds over its citizens (Angwin et al., 2016). The broader role of AI in policing—from facial recognition to predictive algorithms used to allocate police resources—typically meets this description as well (Brayne, 2021). And algorithmic tools process and monitor online speech in order to detect and penalise harmful communication (Gillespie, 2020; Gorwa *et al.*, 2020). With recent advances in generative AI, text-to-image diffusion models have also directly impacted on the prospects of many artists whose work has been used in the training data for those models, and who now find themselves in direct competition with models that can algorithmically generate design products of comparable quality at an infinitesimal fraction of the cost of producing them from scratch. Image and video generation can also be used to create 'deepfakes' that can be used for nefarious purposes from individual blackmail to the spread of disinformation (Rini and Cohen, 2022).

AI is also used to surveil populations—in workplaces by employers using algorithmic management tools (Ajunwa et al., 2017; Ajunwa and Greene, 2019); in society at large by the state—which is a direct harm, and another way in which AI is used to exercise power (Susskind, 2018). Manual surveillance is labour-intensive and expensive. Automated surveillance is neither of those things. Consider how businesses using Office365 can set Natural Language Processing algorithms to monitor employees' communications to detect non-compliant behaviour—a practice that ubiquitous LLMs will only turbocharge.[1] Or how face recognition enables vast CCTV networks to be operationalised to enable near-universal tracking of a population.[2] Or how mass surveillance and big data analytics have been used by ICE agents to hunt undocumented migrants in the US (Bedoya, 2020).

AI and related computational systems also shape people's choices by intervening on their options. For example, they can directly determine people's choice sets, making some options available or unavailable. This is particularly common in our digital lives, where computational systems dynamically adapt to make some things possible, others impossible. This is described by Roger Brownsword as 'technological management'—the practice of shaping people's behaviour not by penalties, but by making undesired behaviour technologically impossible (Lessig, 2006; Brownsword, 2015). This is especially prominent in the use of computational systems to enforce copyright (determining, for example, precisely how you are able to use digital content that you have purchased, or else identifying and preventing the sharing of copyrighted content [Gorwa *et al.*, 2020]). Of course, computational systems using AI can also obstruct dispreferred options, or make them harder to perform, as with the 'dark patterns' that make controlling one's privacy online so challenging (Susser and Grimaldi, 2021). And they can add penalties to dispreferred options, as for example with 'smart contracts' that automatically execute penalty clauses when one party fails to comply. As OpenAI and others have generative AI models like DALL-E 2 and ChatGPT, millions of users worldwide have experienced this technological management very explicitly, as ambiguous terms of service are applied automatically without recourse—these models will refuse to respond to prompts deemed non-compliant, removing the option of, for example, generating an image using a celebrity's face, or preventing the creation of NSFW images. OpenAI and others argue that their efforts to make these agents safe are merely attempts to align the models to human values. This is in part true, but it also results in using the model to govern users.

As well as dynamically removing options, computational tools can dynamically *create* options. This too is a kind of power, roughly analogous to what French philosopher Michel Foucault called 'governmentality' (Foucault, 2010; Bucher, 2018). For example, consider the simple practice of suggesting people to 'friend' or 'follow' on social media. This can create opportunities for new kinds of social interactions (Bucher, 2013). Or consider the personalized delivery of advertisements. Though sometimes a pure nuisance, these ads may involve presenting people with opportunities or options that are not otherwise easily accessible: for example, an investment opportunity, or a job posting, or even drawing the user's attention to a new area where they might consider purchasing a home (Imana et al., 2021). At the intersection of both examples, consider the prospective role of AI in developing the 'metaverse', at the intersection of online and offline lives. Companies like Meta are now establishing the options that will be available to us in this new domain, and AI and related tools will shape what is possible for us within it—making some things impossible, others possible. For example, in its 'Horizons' virtual world, Meta has introduced a four-foot boundary around avatars so that non-consensual invasion of personal space (including virtual sexual assault) is removed as an option.[3] With the advent of generative AI, we are also seeing AI offer new capacities in both text- and image-generation, opening up entirely new sectors of economic activity (while also foreclosing others, as noted above). The range of new options enabled by LLMs is mind-boggling. Even if only 10% of the experiments in how to integrate these systems into our lives bear fruit, they will radically expand the scope of our agency.

Besides actually creating new options, AI and related technologies have given 'nudge' economics a profound shot in the arm, enabling (sometimes crude) ideas from behavioural economics to be operationalised at massive scale, with the ability to do massive social experiments enabling persistent fine-tuning (Thaler and Sunstein, 2008; Kramer et al., 2014). AI is first used to profile individuals in order to tailor particular messages to them, then to deliver the right message for that person (as well as to manage the auction where that message competes with others for their attention), and then to learn from that trial in order to refine the message or select among alternatives (Benn and Lazar, 2021). Karen Yeung has described this as 'hypernudging', but it is really more of a shove than a nudge, and has spawned an entire new field of 'persuasive technology' (Yeung, 2017).

　　Nudging (hyper or otherwise) is supposed to focus mostly on how people's options are presented to them. But hyper-personalised computational systems are also well suited to more directly shaping people's beliefs and desires. Sometimes this is transparent and well-intentioned. At other times it is both deceptive and extractive. In either case, it involves the exercise of power. We can distinguish roughly between three modalities: when AI is used to connect us to (hopefully) authoritative sources of information; when it mediates horizontal communication between users of the internet; and when it is used to identify and target us for specific persuasive messaging. In each case AI systems shape our beliefs and desires, and the people who design these systems thereby exercise significant power over those affected by them. Indeed, search and recommender algorithms deploy, and have driven research in, some of the most advanced techniques in AI, using deep neural networks, large language models, and reinforcement learning among others. One very likely commercial application of the most advanced LLMs is (depressingly enough) in precisely this area, as they are used to generate compelling marketing text. And we can already see that when people use language models to write, if the models are politically 'opinionated', that ends up shaping the users' beliefs too (Jakesch et al., 2023).

　　Consider AI's role in shaping what people know by giving access to (hopefully) authoritative sources of information—for example in helping find relevant public health information during the COVID-19 pandemic. Many people are practically dependent on digital platforms to guide them to this information. The platforms' intentions during the pandemic seemed to be broadly good—they wanted to help people find authoritative sources—but they still had to make many controversial choices about what to show, what to exclude, and what to prioritize—and the 2022 change of leadership at Twitter demonstrated that intentions can easily change with a change of ownership. In any case, there is no 'neutral' path, especially given deep and persistent disagreement among people as to what information matters. They cannot avoid exercising power. And given the structure of the digital information environment, as well as the volume of information available, they have to build these decisions into the search algorithms on which people rely to navigate the functionally infinite internet. If LLMs, or their near future descendants, become our general-purpose interface to all digital technologies—as many are betting they will—then this power will be rendered still more acute, as AI assistants will select, distil, and narrate information for us, mediating communication in explicit ways that inescapably shape what we come to know.

Recommender systems mediate horizontal communication among internet users—determining whose speech is removed, muted, published or amplified. Although human curation and content moderation plays a significant role, substantial automation is unavoidable given the sheer volume of information at stake (Andrejevic, 2013; Roberts, 2019; Gorwa *et al.*, 2020). These systems often rely on algorithms that can optimise for some measurable feature, which may only be an inadequate proxy for the properties that really matter. Engineers must choose not only what to aim at—what kind of information economy they want to achieve—but also how to operationalise that objective by choosing some measurable objective function. And the stakes are high—researchers (e.g. Vaidhyanathan, 2018) have long argued that optimising for user engagement generates adverse social impacts, and we now know that this has also long been recognised by internal researchers at Meta. Prioritising content that generates user engagement has led to the spread of radicalising misinformation, shaping people's beliefs and desires in deeply harmful ways. LLMs risk supercharging the spread of misinformation, as they make it cheap and easy to generate vast amounts of misleading content, as well as highly realistic and interactive deception over time (Bender et al., 2021; Horvitz, 2022).

As well as connecting users to authoritative sources and to each other, the same basic tools are used to connect businesses to users, in order to extract value from the latter. The goal here is clearly on the borderline between persuasion and manipulation (Kaptein and Eckles, 2010; Kosinski et al., 2013; Susser et al., 2019). Online behavioural advertising is sometimes perfectly transparent and non-deceptive—it's simply about showing you a product that you might be interested in, given contextual cues from your entry in a search engine or the website you are visiting. But often it is not. You are being shown this advertisement because a profile of you has been built up from digital breadcrumbs left across multiple apps and websites, as well as edge computing devices (Turow, 2011; Zuboff, 2019). You are receiving *this version* of the advertisement because automated testing over massive populations (none of whom knew they were test subjects) discovered that this wording worked best for people like you. In the extreme scenario, your personality type or your susceptibility to a particular persuasion technique is being automatically operationalized in order to increase the chance that you will be persuaded. And while all of these different measures might make relatively little difference to the probability that *you* will buy the advertised item, over the whole population it does make a difference. Even if persuasive technologies are relatively unsuccessful at manipulating individuals, they enable an accelerated version of 'stochastic manipulation' of populations at large (Benn and Lazar, 2021).

How we see the world is profoundly mediated by our digital infrastructure, and AI is now integral to that infrastructure. Choices made by the designers, developers, and deployers of AI systems determine how the world is represented to us. Even were these decisions not made with any particular intention to shape people's behaviour one way or another, the ability to structure how billions of people perceive the world, through search algorithms and recommender systems, involves an extraordinary level of power. This will only increase as, instead of relying on AI to direct us to primary sources online, we increasingly depend on AI assistants to curate, create, edit, and dynamically summarise that content.

In each of these domains, Automatic Authorities (like most technologies) are 'force multipliers'. They enable fewer people to achieve bigger impacts on a wider range of choices in the lives of more people. They increase the degree, the scope, and the concentration of power at stake.

For example, by automating the application of penalties, Automatic Authorities increase the probability that non-compliance will be penalised. This increases the deliberative weight of that penalty as surely as increasing the severity of the penalty. Consider 'smart contracts': if you miss a payment on your car lease, instead of triggering an unreliable process of debt collection, the vehicle will simply be automatically disabled until payments resume (Susskind, 2018; Zuboff, 2019). This increases the bank's degree of power over the debtor.

These amplifications of the abilities of the powerful to affect the lives of those subject to their decisions also increase the *scope* of their power. Decisions that would, in the past, have been up to B alone now become susceptible to interference, in part because of our growing dependence on computational systems in ever more spheres of our lives. The more we rely on them, the greater the scope of choices that Automatic Authorities can influence.

Automatic Authorities also make the exercise of power more *efficient*, and therefore more *concentrated*. They operate at massive scale, enabling one individual to affect millions, even billions of lives, as when Mark Zuckerberg or Elon Musk decides that Facebook or Twitter respectively should adjust their feed algorithms. Relatedly, the increasing role of computational systems in public governance concentrates power in the hands of CEOs of private companies. Take Northpointe's COMPAS recidivism prediction algorithm (Angwin *et al.*, 2016). In the past, no individual could influence bail decisions across multiple jurisdictions in the US except through the proper legislative and judicial processes. But the CEO of Northpointe can influence *many* such decisions; their choice to focus on one understanding of fairness rather than another (for example) ramifies across thousands of cases (Hellman and Creel, 2021). AI concentrates power in fewer hands. And as AI systems advance, if we become increasingly reliant on vast and inscrutable foundation models, power will increasingly be concentrated in the hands of those systems themselves, more so than the people who deploy them. This raises profound and serious normative challenges, which societies must urgently address.

**Justifying Power**

Automatic Authorities are used to exercise power. They have enabled new power relations, and intensified other ones. And they allow significant concentration of power. But why does this matter? In particular, what distinctive normative questions are raised by invoking *power*? After all, if AI affects people's interests, their choices, their beliefs and their desires, couldn't we simply evaluate all of those effects against, say, a principle of distributive justice? Why not say that we should use AI systems to shape people's lives in ways that, for example, make the worst off group in society as well-off as they can feasibly be? Or why not just try to maximise the benefits and minimise the costs, and choose the path with the best (perhaps weighted) sum of the two? Or, indeed, why not aim to use AI systems to achieve the goals described in one of the (many) lists of 'AI Ethics' principles? Why can't we simply apply one of these standards of *substantive justification* to the use of AI to exercise power?

    The answer: the exercise of power by some over others generates presumptive moral objections grounded in individual freedom, social equality, and collective self-determination, which can be answered only if power is used not only for good ends, but legitimately and with proper authority.

    The first step is to show that the exercise of power generates presumptive objections, independent of what it is used for. Start with the objection from individual freedom. On a simple, negative conception, one's freedom consists in the absence of external interference in one's choices. A more complex conception would also emphasise the absence of the *risk* of such interference (Carter, 2008; Kramer, 2008). A further, republican, extension, would add emphasis on reducing the *possibility* of arbitrary interference (Pettit, 1997). Positive theories of freedom generally add that one must not only avoid interference, but also have an adequate range of options, and the capacities and information necessary to choose wisely among them, based on one's authentic desires (Raz, 1986). As described in the previous section, AI can be used to directly limit people's negative freedom—to incarcerate and surveil them—as well as to limit their options, and indeed to shape their ability to act authentically to fulfil their desires based on accurate beliefs.

    Of course, AI is also used in ways that enhance people's individual

freedom—power and freedom are not strict duals. There is a somewhat deeper tension between power and social equality. Social equality is the existence of social relations whereby people treat one another as equals, and this equal treatment is reflected in their institutions (Anderson, 1999; Kolodny, 2014). Although in many respects—esteem, affection, and so on—we are *not* equal even in egalitarian societies, in one fundamental and important sense each citizen is the equal of every other. We have the same basic rights and duties, the same standing to invoke the institutions of the state, the same opportunity to participate in them, the same ability to contribute to setting the shared terms of our social existence. The power of some over others involves a hierarchical one-way relationship in which A exercises power *over* B, leaving the two in unequal social relations—irrespective of whether A treats B well or poorly. This is a particular concern for the exercise of power by means of AI, both because it relies on expert knowledge that is far beyond the ken of most of those subject to it, and because much of the power exercised by means of AI structures digital environments in which we have long abandoned any pretence of social equality, substituting the aspirationally egalitarian liberal democracies of our offline lives for digital feudalism, subject to the whims of a tiny handful of unaccountable executives.

Social equality can be satisfied if people have an equal opportunity to shape the shared terms of their social existence, even if they do not actively take up that opportunity. Collective self-determination is to social equality much as positive freedom is to negative freedom: it is about (enough) people actually positively shaping the world in accordance with their values. If the power of some to shape the shared terms of others' social existence is not an expression (in some sense) of the collective's will, it is presumptively antithetical to collective self-determination. At present, society's collective dependence on the whims of executives leading 'Big Tech' companies undermines collective self-determination, as well as social equality.

Even if AI is used to exercise power for goals that serve freedom, equality, and self-determination, or other equally important values, these pro tanto objections focus not strictly on what power is being used to achieve, but rather on the fact that power is being exercised at all. As such, even if AI is used to exercise power for noble ends, these presumptive objections still apply. They *might* be overridden by the good being done, but they can be fully resolved only if power is exercised not only for the right ends, but *in the right way*, and *by the right people*. These are the standards of *substantive justification*, *procedural legitimacy*, and *proper authority*, or, more colloquially, the 'what', 'how' and 'who' standards.

The standard of substantive justification simply demands that power is used to achieve justified ends. This applies whether one invokes power or not—it aims at fair outcomes, the promotion of well-being and autonomy, and the many other goods that modern liberal democracies typically aim at.

Whenever a new technology generates new capacities, it raises novel questions of substantive justification. For example, AI now enables a level of automated influence far in excess of what we could achieve with earlier technologies. When is such influence morally permissible, and when does it constitute privacy-invasive or otherwise exploitative manipulation (Benn and Lazar, 2021)? Through algorithmic online media, we have a novel capacity to distribute collective attention at scale across vast populations. What moral principles should govern such practices? The advent of AI calls for us to develop new theories of communicative justice (Lazar, 2022). The increasingly pervasive use of AI-based predictive technologies invites us to ask whether predictive models themselves can be just or unjust, or whether instead we should design predictive models with outcome justice squarely and uniquely in mind (Lazar and Stone, 2022). Generative AI similarly raises many new questions of substantive justification: for example, if a model were trained only on images that were legitimately in the public domain, but is able to cheaply produce art that is commercially competitive with the work of professional artists, do the latter have a justified complaint against the widespread dissemination of such a model?

On the standard of procedural legitimacy, it's not enough to use power to achieve justified ends; it must also be exercised in the right way, by following appropriate procedures. For the exercise of power to be consistent with individual freedom and social equality, it has to be subject to strict constraints. We preserve our freedom, and ensure that we collectively have power over those who individually have power over us, by limiting their power. The nature of these limits depends on the kind of power being exercised. We can get some insight into them by thinking about the core standards of the rule of law: as well as getting matters substantively right, the governing agent should be consistent, and (morally) treat like cases alike; those subject to the decision should have the opportunity and ability to understand why the decision has been made; standards of due process should be met where feasible, and those exercising power should be subject both to processes of contestation by those subject to their decisions, and to review and potential dismissal on discovery of persistent misconduct (Waldron, 2011). Not all cases of the exercise of power by AI are subject to such exacting standards—but AI is often used to govern, whether by states, by private companies on behalf of states, or by private companies governing domains of life that states have left functionally ungoverned. Governing involves setting and enforcing rules, and in particular shaping power relations among the governed (Lazar, 2023). In these cases, the exercise of power by means of AI has fallen pretty far short of all of these standards. Indeed, one might think that when we use highly complex and inscrutable ML-based techniques to implement and enforce rules, we are constitutively prevented from meeting these kinds of procedural standards. This is one important foundation for widespread concern about the need for AI systems to be explainable—when they are used to exercise power, especially governing power, certain kinds of explanations will be essential for them to exercise power legitimately (Lazar, 2024).

On the third, authority, standard it matters not only that power is being exercised for the right purposes and in the right way, but also by the right people. The people exercising power should be those with the authority to do so within that institution. If those who exercise power around here lack authority to do so, then we cannot be collectively self-determining. The criteria for proper authority vary depending on the nature of the institution, but a key general point is that the more pervasive and important an institution is in the lives of a group of people, the more prima facie important it is that the authority to govern it should stem from them, the people served by that institution. Our information, material, creative and other economies are significant parts of our lives—they are important. They are also increasingly reliant on digital platforms, which are themselves structured by AI, in particular recommender and search algorithms. And those algorithms are designed and implemented by a small number of employees of private businesses which lack any suitable authority to so extensively shape the shared terms of our social existence. Unauthorised power is a threat to our collective self-determination.

To illustrate this point, consider the controversy over the use of smartphones to support COVID-19 contact tracing. They offered the promise of identifying anonymous or forgotten close contacts during a COVID patient's asymptomatic infectious period. But they seemed likely to be most effective where they could be fully integrated with manual contact tracing. This would come, however, at a cost to individual privacy, since it could create a centralised database of people's close contacts. This is a tricky trade-off to navigate. However, the only way to get these apps to work successfully on Android and iOS was to work with the Bluetooth protocol developed by Google and Apple for this purpose. And Google and Apple decided, for all of us, how to weigh privacy against public health: they adopted a decentralised exposure notification protocol that could not augment manual contact tracing. Setting aside whether they weighed these values correctly, it is deeply counterintuitive to suggest that *they*, rather than the public health authorities of liberal democracies, had the authority to make that decision. This seems obviously in tension with collective self-determination, as well as the principles of social equality that underpin democratic decision-making.

**Conclusion**

Power over is the social relation where an agent A can significantly affect the interests, options, beliefs and desires of another agent B but not vice versa. AI and related computational systems are being used by some to exercise power over others. They enable new and intensified power relations, and a greater concentration of power. This is especially clear in our online lives, which are increasingly structured and governed by computational systems using some of the most advanced techniques in AI. But it is also apparent in our offline lives, as computational systems using AI are used by powerful actors including states, local government, and employers. We are everywhere subject to 'Automatic Authorities'—automated systems whose power over us we have automatically accepted. Recent breakthroughs in generative AI, especially large, self-supervised models using transformers have induced a very real prospect of AI systems themselves exercising significant power over us, as they provide a general interface to the internet, and so promise to mediate (and therefore substantially shape) our access to information and communication.

Proponents of various principles of 'AI Ethics' sometimes imply that the sole normative function of those principles is to ensure that AI is used to achieve socially acceptable goals. They imply that substantive justification is sufficient for all-things-considered justification of these uses of AI. This is especially true of those who focus on 'aligning' LLMs, or making them 'safe'. Drawing attention to the ways in which AI systems are used to exercise power demonstrates the inadequacy of this normative analysis. When new and intensified power relations develop, we must attend not only to what power is used for, but also to how and by whom it is used: we must meet standards of procedural legitimacy and proper authority, as well as substantive justification.

This has pressing practical implications for those who design and deploy existing AI systems. Many of those who do so have already started to ask themselves crucial questions of substantive justification—recognising that the tools they deploy can have better or worse social impacts, can conduce to justice or injustice, and aiming for positive impact. But this is emphatically not enough. Even when these tools are used to achieve broadly good ends, when they are used to exercise power they must also be used in the right ways, and crucial decisions must be taken by those with proper authority to do so. Those engaged in developing and deploying AI technologies must reflect on the power that their control over these technologies vest in them—are they using that power in appropriate ways? And perhaps most importantly, should they have that power in the first place? This is not, importantly, an invitation only to engage in methods of participatory design that draw on inputs from affected populations in order to shape Automatic Authorities (Harrington et al., 2019). That is important, but is not sufficient, as long as the ultimate decision about how the system will operate sits with the company deploying it, rather than with the people themselves, preferably through the auspices of democratic institutions.

The 'how' and 'who' standards do not always apply with equal force. They matter most when the stakes are high, and when power is used for the purposes of governance. But with AI systems increasingly being used in high-stakes applications that involve governing those affected by them, the status quo, in which groups of researchers and engineers at private, for-profit companies decide how vast populations worldwide will be governed, is unsustainable.

In the near future, it seems likely that LLMs and their descendants will play the role that recommender systems and search currently play in shaping our information and communications environments. Indeed, some aspire to a situation where AI systems govern vast swathes of human activity, not just information and communication (themselves vast enough). These models are already sufficiently complex and inscrutable that, if deployed at this kind of scale, they might not be under the effective control of those who design and deploy them. If so, they will effectively be exercising power themselves. In recent years, the field of 'AI Safety' has emerged, with one of its principal goals being to ensure that such an outcome is consistent with ongoing human flourishing (Amodei et al., 2016).

This research and political agenda might be predicated on an unrealistic assessment of technological feasibility (Bender *et al.*, 2021). Perhaps LLMs will continue to be plagued by 'hallucination', their tendency to produce authoritative but false answers that non-experts cannot distinguish from the truth (Tam et al., 2022). And perhaps they will unavoidably continue to reproduce the prejudices encoded in their training data (Weidinger et al., 2021). But suppose those challenges could be overcome, as well as the other technical challenges more directly targeted by AI Safety research (Hendrycks et al., 2021). The deeper problem is that this approach focuses too narrowly on answering the 'what' question of substantive justification. Ensuring that powerful AI systems are 'provably beneficial' (Russell, 2019) or 'value-aligned' is insufficient for their widespread deployment in high stakes applications to be justified. If they are going to exercise power—as they surely will in these scenarios—the deeper question is whether they can ever do so with proper authority, in ways that are procedurally legitimate. As well as the obvious problems raised for any publicity requirements on legitimacy of our depending on vast inscrutable models that are at best only partially understood, it seems extremely unlikely that an Automatic Authority could ever *itself* be a justified authority. For example, at a first pass it seems extremely unlikely that such a system would be consistent with collective self-determination at least, and one might reasonably question whether a society that hands over self-governance in important domains to computer models is really a society of equals. And given our well-known inability to predict precisely how these models will behave, such a transfer of power would seem obviously to undermine our liberty, placing us at significant risk of wrongful limitations on our freedom.

As a result, we should be very sceptical about whether it could ever be all-things-considered appropriate to vest this kind of power in Automatic Authorities, however sophisticated or aligned. This has significant implications in the present: researchers and engineers currently engaged in research predicated on the assumption that we will, at some stage in the future, entrust AI systems with governing power have compelling reasons to rethink their research direction. The goal should not be to ensure that all-powerful AI systems are reliably aligned with human interests, but rather to ensure that we prevent the deployment of any such AI systems in the first place. Even if they could in theory govern justifiably (which is itself very unlikely on current evidence), if they are remotely similar to existing ML systems, they very likely could not do so legitimately and with proper authority. Until we have resolved those—perhaps unsolvable—challenges, we should no more invite their arrival than we should aspire to hand over the reins of power to a benevolent but capricious human dictator.

---

[1] https://www.microsoft.com/en-us/microsoft-365/blog/2020/02/20/leverage-ai-machine-learning-address-insider-risks/.

[2] https://www.cnbc.com/2019/05/16/this-chinese-facial-recognition-start-up-can-id-a-person-in-seconds.html.

[3] https://www.businessinsider.com/meta-metaverse-virtual-groping-personal-boundary-safety-bubble-horizons-venues-2022-2.